\documentclass[aps,prl,twocolumn,groupedaddress,showpacs]{revtex4}

\usepackage{graphicx}
\begin{document}

\title{Resonant Tunneling in Truly Axial Symmetry Mn$_{12}$ 
Single-Molecule Magnets: Sharp Crossover between 
Thermally Assisted and Pure Quantum Tunneling}


\author{W. Wernsdorfer$^1$, M. Murugesu$^2$, and G. Christou$^2$}


\affiliation{
$^1$Lab. L. N\'eel, associ\'e \`a l'UJF, CNRS, BP 166,
38042 Grenoble Cedex 9, France\\
$^2$Dept. of Chemistry, Univ. of Florida,
Gainesville, Florida 32611-7200, USA
}

\date{\today}

\begin{abstract}
Magnetization measurements of a truly axial symmetry
Mn$_{12}$-$t$BuAc molecular nanomagnet 
with a spin ground state of $S = 10$ show
resonance tunneling. 
This compound has the same magnetic anisotropy
as Mn$_{12}$-Ac but the molecules are better isolated and
the crystals have less disorder and a higher symmetry.
Hysteresis loop measurements at several temperatures reveal
a well-resolved step fine-structure which is due to level crossings
of excited states. All step positions can be
modeled by a simple spin Hamiltonian. The
crossover between thermally assisted and pure quantum 
tunneling can be investigated with unprecedented detail.
\end{abstract}


\maketitle

${\rm [Mn_{12}O_{12}(O_2CCH_3)_{16}(H_2O)_4]\cdot 2CH_3CO_2H\cdot 4H_2O}$, 
Mn$_{12}$-Ac for short, 
was the first single-molecule magnet (SMM), 
exhibiting slow magnetization relaxation of its $S$ = 10 spin ground state 
which is split by axial zero-field splitting~\cite{Sessoli93}. 
It was also the first system 
to show thermally assisted tunneling of magnetization
~\cite{Novak95,Friedman96,Thomas96}.
During the last several years, many more SMMs
have been discovered and they are now
among the most promising candidates for observing 
the limits between classical and 
quantum physics since they have a well defined 
structure, spin ground state 
and magnetic anisotropy~\cite{Sangregorio97,Aubin98,Caneschi99,Yoo_Jae00}.
Nevertheless, Mn$_{12}$-Ac is still the most widely studied 
SMM~\cite{Chudnovsky97,Prokofev98,Kent00,Kent00b,ChiorescuMn12PRL00,Mertes01,Garanin02,Cornia02,Dressel03,Bal04b,Takahashi04,Park04a,Luis04}.
While a rough understanding of the quantum phenomena in
Mn$_{12}$-Ac was clear from the early studies, 
a detailed understanding has not yet emerged.
For example, current theoretical models
assume the presence of quadratic and quartic transverse
crystal-field interactions in the spin Hamiltonian, 
where the former has been ascribed
to solvent disorder~\cite{Cornia02}. However, these interactions,
which contain only even powers of the raising and lowering
operators, do not provide an explanation for the observation
of odd tunneling steps in the hysteresis loops. 
It has been proposed that easy-axis tilting 
might give the missing odd transverse interactions~\cite{Takahashi04}.
Although such solvent disorder induced tilts exist,
the tilt values are still unclear~\cite{WW_PRB04}.
Hyperfine, dipolar, and Dzyaloshinsky-Moriya interactions
were also proposed to
be responsible for odd transverse terms~\cite{Prokofev98,ChiorescuMn12PRL00}.

Other theoretical and experimental studies concern the
crossover between thermally assisted and pure quantum 
tunneling~\cite{Chudnovsky97,Kim,Kent00,Kent00b,ChiorescuMn12PRL00,Mertes01,Garanin02}.
The crossover can occur in a narrow temperature interval with
the field at which the system crosses the anisotropy
barrier shifting abruptly with temperature, 
or the crossover can occur in a broad interval
of temperature with this field changing 
smoothly~\cite{Chudnovsky97,Kim,Garanin02}.
The first studies were published
on Mn$_{12}$-Ac~\cite{Kent00,Kent00b,ChiorescuMn12PRL00}
but significant distributions of molcular environments 
are unfortunately present, and these complicate
the interpretation of the data~\cite{Mertes01,Cornia02}.

We present here resonant quantum tunneling measurements 
of a recently discovered analog of Mn$_{12}$-Ac, namely  
${\rm [Mn_{12}O_{12}(O_2CCH_2Bu^{\it t})_{16}(CH_3OH)_4]\cdot CH_3OH}$, 
called Mn$_{12}$-$t$BuAc henceforth.
We show that this compound has the same magnetic anisotropy
as Mn$_{12}$Ac but the molecules are better isolated
and the crystals contain less disorder and a higher symmetry.
Hysteresis loop measurements at several temperatures reveal
a fine structure of steps which is due to the dominating
energy level crossings. All step positions can be
modeled by a simple spin Hamiltonian. The
crossover between thermally assisted and pure quantum 
tunneling is investigated.

\begin{figure}
\includegraphics[width=.50\textwidth]{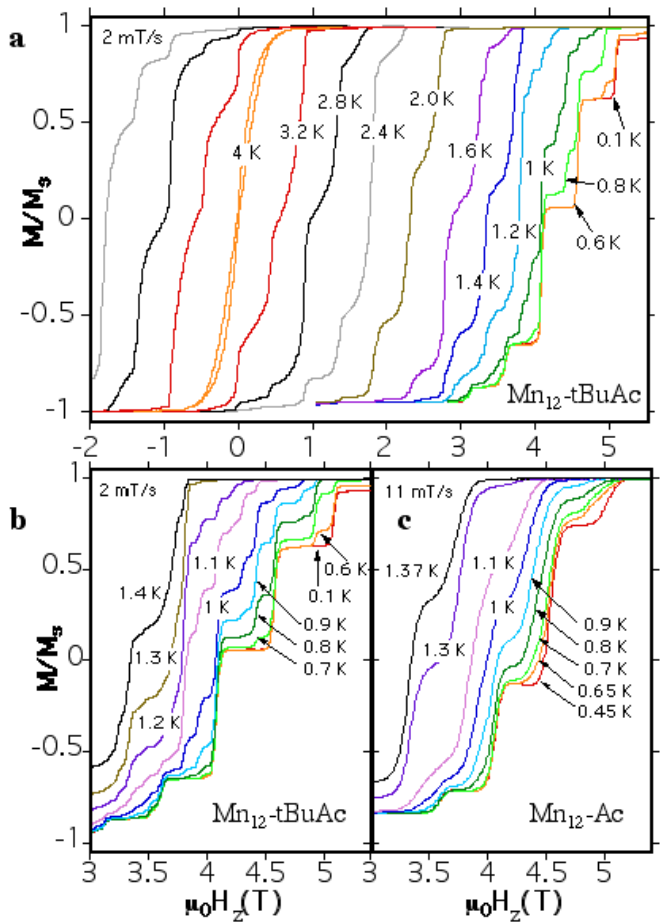}
\caption{(color online) Hysteresis loops of single crystals 
of (a-b) Mn$_{12}$-$t$BuAc and (c) Mn$_{12}$-Ac molecular clusters
at different temperatures and a constant field
sweep rate indicated in the figure. The data in (c) were
taken from~\cite{ChiorescuMn12PRL00}.
The loops display a series of steps, 
separated by plateaus. As the temperature is lowered, there is a decrease in 
the transition rate due to reduced thermal assisted tunneling. The hysteresis 
loops become temperature independent below 0.6~K, demonstrating 
quantum tunneling at the lowest energy levels.}
\label{hyst_Mn12ac_tBuAc}
\end{figure}

${\rm [Mn_{12}O_{12}(O_2CCH_2Bu^{\it t})_{16}(H_2O)_4]}$
was prepared by the carboxylate substitution procedure 
described elsewhere~\cite{Soler01}, and crystallizes 
in triclinic space group $P\bar{1}$~\cite{Soler03b}. 
However, recrystallization from mixed CH$_3$OH/Et$_2$O 
solvent gave Mn$_{12}$-$t$BuAc in tetragonal space 
group $I\bar{4}$. Full details of the synthesis, 
crystal structure and magnetic characterization 
are presented elsewhere~\cite{Murugesu05}; the ground 
state spin $S = 10$ was established by magnetization 
measurements in the 1 - 7 T and 1.8 - 4.0 K ranges. 
The molecular structure of Mn$_{12}$-$t$BuAc is very 
similar to that of Mn$_{12}$-Ac except that the 
acetate (Ac) and H$_2$O groups have been replaced 
by Bu$^t$CH$_2$CO$_2$ ($t$BuAc) and CH$_3$OH groups, 
respectively. The increased bulk of the Bu$^t$CH$_2$CO$_2$ 
groups leads to a unit cell volume for Mn$_{12}$-$t$BuAc 
(7.06 nm$^3$) that is almost double that of 
Mn$_{12}$-Ac (3.72 nm$^3$), and thus to greater 
intermolecular separations and decreased intermolecular 
interactions relative to Mn$_{12}$-Ac. In addition, 
the interstitial CH$_3$OH solvent molecules 
in Mn$_{12}$-$t$BuAc are not disordered and neither 
are they hydrogen-bonding with the Mn$_{12}$ molecules. 
As a result, the site symmetry of the latter in Mn$_{12}$-$t$BuAc
is truly axial with a small distribution of environments. 
This is in stark contrast to Mn$_{12}$-Ac where each 
of the acetic acid (CH$_3$CO$_2$H) molecules in the 
crystal forms a strong OH$\cdot\cdot\cdot$H 
hydrogen-bond with a Mn$_{12}$ molecule but will 
do so with only one of the two Mn$_{12}$ molecules next to it. 
Since each Mn$_{12}$ molecule is surrounded by four 
CH$_3$CO$_2$H molecules, this disorder in the acetic 
acid orientation leads to the Mn$_{12}$ molecules in 
Mn$_{12}$-Ac being hydrogen-bonded with $k$ 
CH$_3$CO$_2$H molecules ($k = 0 - 4$), with the $k = 2$ situation 
also having two possibilities (the two CH$_3$CO$_2$H
attached $cis$ (adjacent) or $trans$ (opposite) about 
the Mn$_{12}$ molecule). Thus, although the Mn$_{12}$-Ac 
crystal possesses a crystallographic average symmetry 
of $I\bar{4}$, it contains a mixture of Mn$_{12}$ molecules 
in six different hydrogen-bonded forms~\cite{Cornia02}, 
and only two of these forms possess axial $S_4$ site symmetry, 
the $k = 0$ and 4 forms. The other forms have lower 
(rhombic) symmetry. As a result, Mn$_{12}$-Ac crystals 
contain Mn$_{12}$ molecules with a wide distribution 
of environments. Consequently, crystals of Mn$_{12}$-$t$BuAc 
are concluded to be far superior to those of Mn$_{12}$-Ac 
for detailed studies such as those in this paper, and 
this has indeed turned out to be the case. Finally, 
the axes of the tetragonal site symmetry of each molecule 
coincide with those of the unit cell, which is not the 
case for Mn$_{12}$-Ac 

The magnetization measurements were 
performed by using a magnetometer consisting
of several 6~$\times$~6~$\mu$m$^{2}$ Hall-bars~\cite{Sorace03}
on top of which a single crystal of 
Mn$_{12}$-$t$BuAc was placed.
The field can be applied in any direction by separately 
driving three orthogonal superconducting coils. 
The field was aligned with the easy axis of magnetization using
the transverse field method~\cite{WW_PRB04}.
The applied field $H_z$ was corrected 
because the determination of the resonance fields must
take into account the internal magnetic field~\cite{note1}.
The sample dimensions were about 
20 $\times$ 15 $\times$ 10 $\mu$m. 
The Hall bars were patterned by 
Thales Research and Technology in Orsay, 
using photolithography and dry etching, 
in a delta-doped AlGaAs/InGaAs/GaAs pseudomorphic 
heterostructure grown by Picogiga International.

\begin{figure}
\includegraphics[width=.50\textwidth]{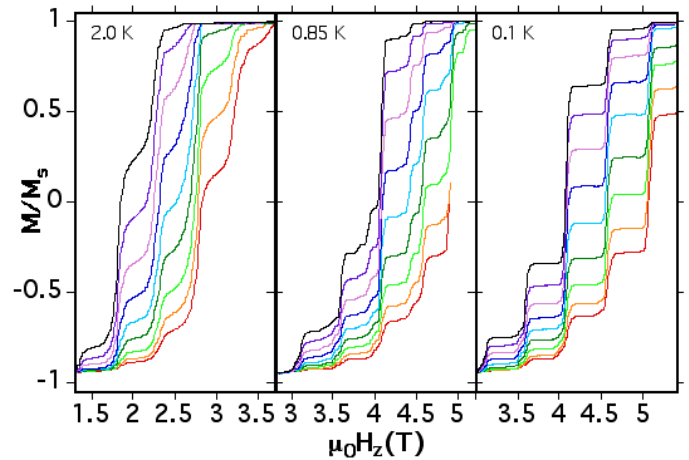}
\caption{(color online) Hysteresis loops of single crystals 
of Mn$_{12}$-$t$BuAc at several field sweep rates and
at 2, 0.85 and 0.1 K. The field sweep rates from
top to bottom are 0.2, 0.5, 1, 2, 4, 8, 17, 35, and 70 mT/s.}
\label{hyst_Mn12tBuAc_v}
\end{figure}

\begin{figure}
\begin{center}
\includegraphics[width=.50\textwidth]{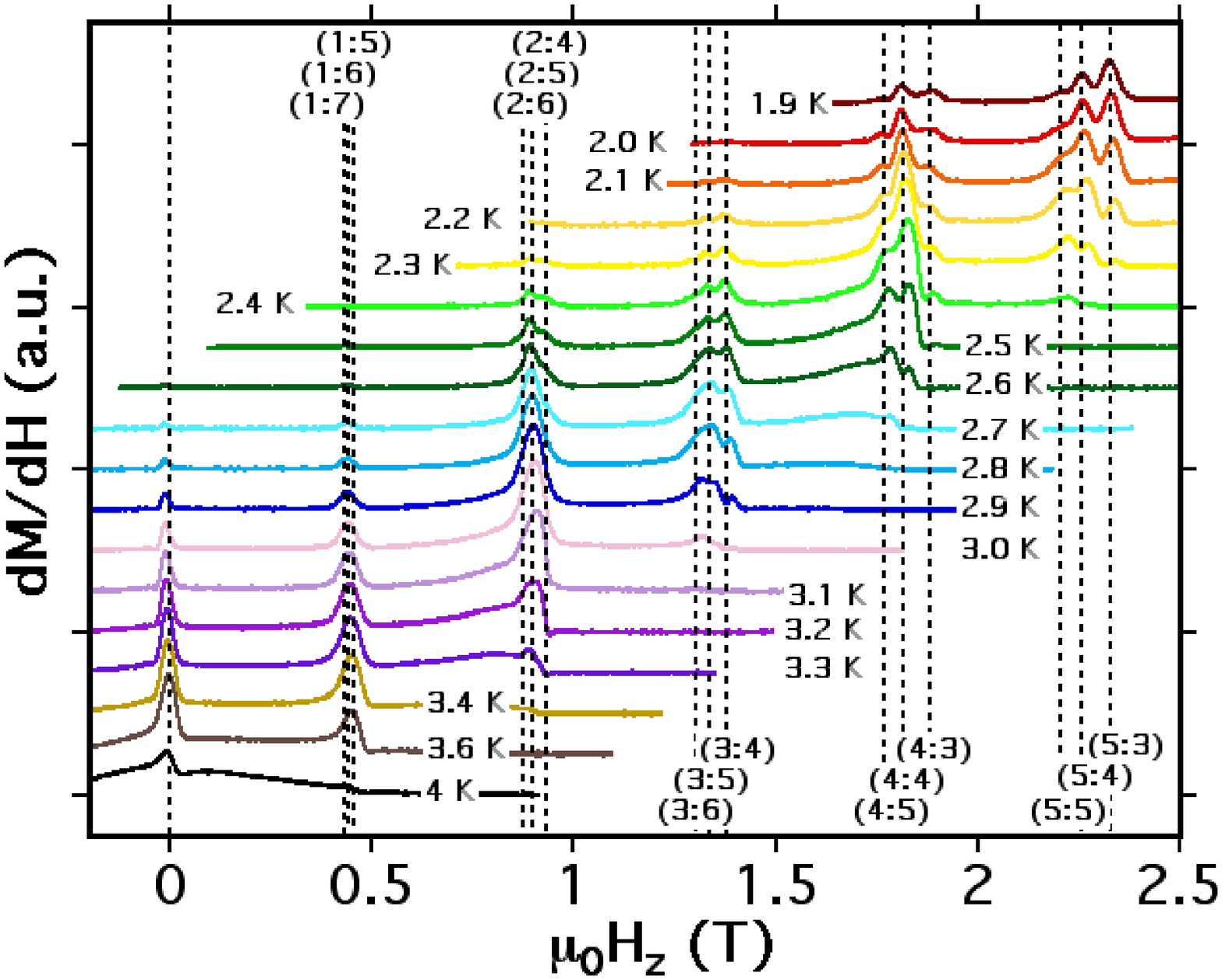}
\includegraphics[width=.50\textwidth]{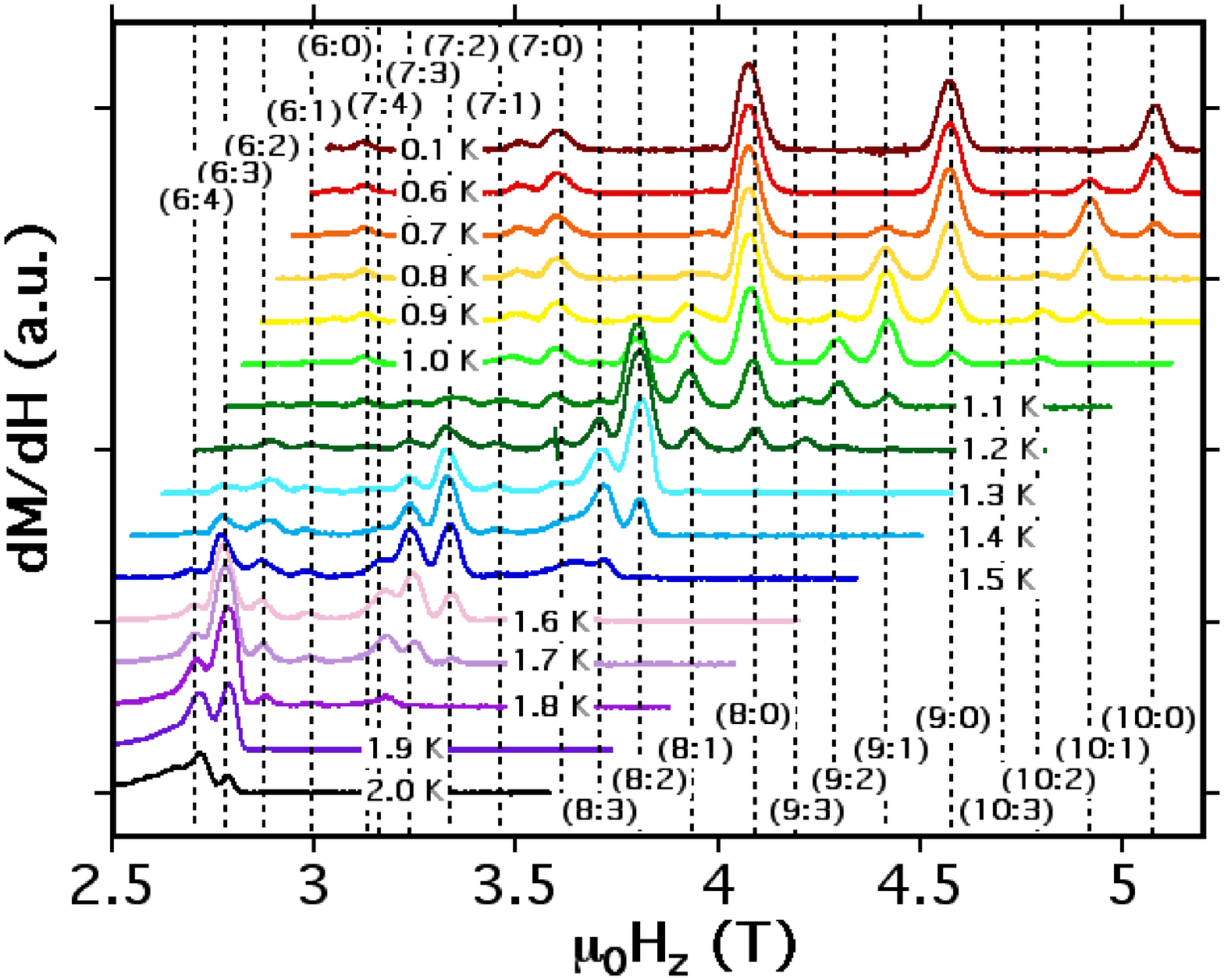}
\caption{(color online) Field derivative of the hysteresis loops of single crystals 
of Mn$_{12}$-$t$BuAc at different temperatures. The applied field was
swept from -6 T to 6 T at a constant field sweep rate of 2 mT/s. 
Resonant quantum tunneling of magnetization occurs at the peaks of $dM/dH$.
The corresponding level crossings are labeled with two indexes $(n:p)$.
The peaks coming from the faster relaxating impurity phase are not labeled.}
\label{dM_dH}
\end{center}
\end{figure}

Fig.~\ref{hyst_Mn12ac_tBuAc} shows the temperature dependence of
the hysteresis loops of Mn$_{12}$-$t$BuAc and Mn$_{12}$-Ac SMMs.
The loops display a series of steps, 
separated by plateaus. 
As the temperature is lowered, the
hysteresis increases because there is a decrease in 
the transition rate of thermal assisted tunneling~\cite{Friedman96,Thomas96}. 
The hysteresis loops become temperature independent below 0.6~K, demonstrating 
quantum tunneling at the lowest energy 
levels~\cite{Kent00,Kent00b,ChiorescuMn12PRL00}.
It is important to note that the loops remain
strongly sweep rate dependent below 0.6K (Fig.~\ref{hyst_Mn12tBuAc_v}).
Apart from the major steps, these hystersis loops 
reveal fine structure in the thermally activated regime
which is also strongly sweep rate dependent (Fig.~\ref{hyst_Mn12tBuAc_v}).
This fine structure was first observed for 
Mn$_{12}$-Ac~\cite{Kent00,Kent00b,ChiorescuMn12PRL00}, see
Fig.~\ref{hyst_Mn12ac_tBuAc}c, but it is much
clearer for Mn$_{12}$-$t$BuAc (Figs.~\ref{hyst_Mn12ac_tBuAc}a, 
\ref{hyst_Mn12ac_tBuAc}b, and~\ref{hyst_Mn12tBuAc_v}).
A convenient way of determining the field positions of the 
steps is to plot the derivative of the magnetization with respect
to the applied field (Fig.~\ref{dM_dH}). The step positions, that
is the maxima of the relaxation rate, are given by the peaks on the 
$dM/dH$ plot.

\begin{figure}
\begin{center}
\includegraphics[width=.50\textwidth]{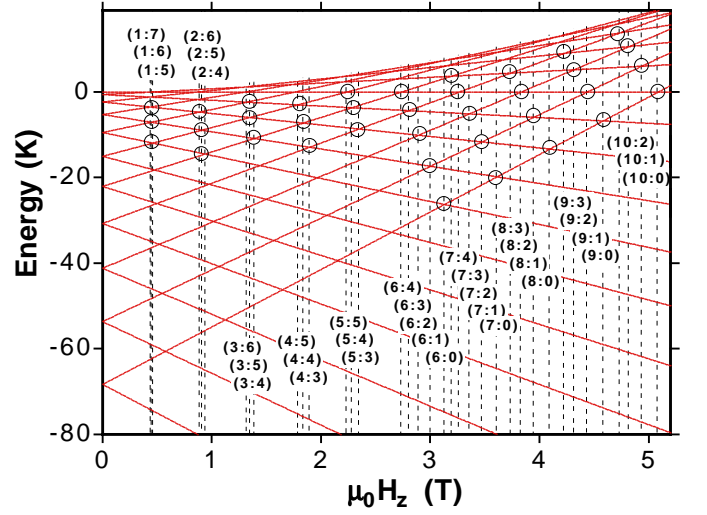}
\caption{(color online) Zeeman diagram of the 21 levels of 
the $S = 10$ manifold of Mn$_{12}$ as a 
function of the field $H_z$ applied along the easy axis. 
At $H_z = 0$, from bottom to top, the levels are labeled 
with quantum numbers $m = \pm10, \pm9, ..., 0$. 
The resonant quantum tunneling steps which lead 
to a step height larger than 0.02 $M_{\rm S}$ 
at 2 mT/s occur 
at the indicated level crossings which are
labeled with two indexes $(n:p)$.}
\label{Zeeman_Mn12}
\end{center}
\end{figure}

The simplest model describing the low-temperature
spin dynamics of Mn$_{12}$-$t$BuAc
has the following spin Hamiltonian
\begin{equation}
	\mathcal{H} = -D S_z^2 - B S_z^4
	- g_z \mu_{\rm B} \mu_0 S_zH_z  + \mathcal{H}_{\rm trans} 
\label{eq_H_biax}
\end{equation}
where $S_x$, $S_y$, and $S_z$ are the three 
components of the spin operator, 
$D$  and $B$ are the anisotropy constants, 
the third term is the Zeeman 
energy associated with an applied field $H_z$,
and the last term ($\mathcal{H}_{\rm trans}$) describes
small transverse terms containing $S_x$ and $S_y$ 
spin operators. Although $\mathcal{H}_{\rm trans}$ produces
tunneling, it can be neglected when determining the field
positions of the level crossing because it is much smaller than the
axial terms. Without $\mathcal{H}_{\rm trans}$, the
Hamiltonian is diagonal and the field dependence
of the energy levels can be calculated analytically (Fig.~\ref{Zeeman_Mn12}).
The energy level spectrum with $(2S+1) = 21$ values 
can be labeled by the quantum numbers $m = -10, -9, ..., 10$. 
At $\vec{H} = 0$, the levels $m = \pm10$ 
have the lowest energy. 
When a field $H_z$ is applied, the energy levels with 
$m < 0$ increase, while those 
with $m > 0$ decrease (Fig.~\ref{Zeeman_Mn12}). Therefore, 
energy levels of positive and negative 
quantum numbers cross at certain fields. 
The field position of the crossing of level 
$m=-S+p$ with $m'=S-n-p$ is given by
\begin{equation}
	H_{(n:p)}=\frac {n\left[D+B\left((-S+p)^2+(S-n-p)^2\right)\right]}
	{g_z \mu_{\rm B}\mu_0 } 
\label{eq_H_step}
\end{equation}
where $n = -(m+m')$ is the step index and $p = S + m$
labels the excited states ($p=0$ for the ground state,
$p=1$ for the first excited state, etc.).

\begin{figure}
\begin{center}
\includegraphics[width=.50\textwidth]{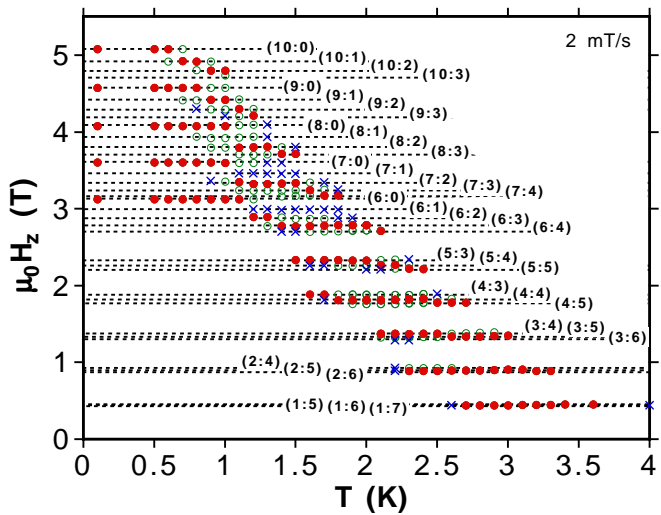}
\caption{(color online) Temperature dependence of the peak positions of $dM/dH$ 
in fig.~\ref{dM_dH} at 2 mT/s. The horizontal lines indicate the calculated
energy level crossing fields.
The largest step for each $n$ are filled dots whereas
the others are open dots or crosses for step heights larger 
or smaller than 0.03 $M_{\rm S}$, respectively.}
\label{H_step}
\end{center}
\end{figure}

\begin{figure}
\begin{center}
\includegraphics[width=.50\textwidth]{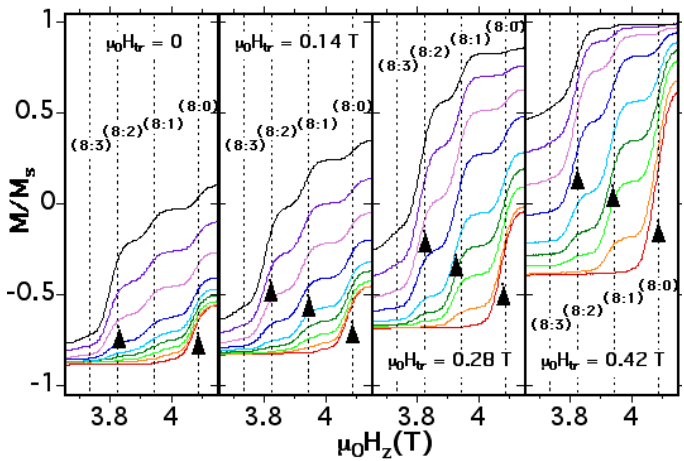}
\caption{(color online) Details of hysteresis loops at 
step $n = 8$ measured in the presence of 
four transverse fields $H_{\rm tr}$ and
at temperatures of 0.1, 0.8, 0.9, 0.95, 1.0, 1.05, 1.1, 1.15, 
and 1.2 K from bottom to top. The field was ramped from -6 T 
to 6 T at a rate of 17 mT/s. The dominating steps are indicated.}
\label{hyst_T_h_tr}
\end{center}
\end{figure}

The step positions $H_{(n:p)}$, determined from Fig.~\ref{dM_dH},
are shown in Fig.~\ref{H_step}~\cite{note2}. The horizontal lines 
indicate the calculated energy level crossing fields
using Eq.~\ref{eq_H_step} with $D$ = 0.563 K, $B$ = 1.2 mK, 
and $g_z$ = 2 where the latter was measured by EPR~\cite{Takahashi04}.
These values are very close to those of Mn$_{12}$-Ac
establishing that the magnetic cores of both molecules
are comparable. Because the resonance fields of all avoided
level crossings are well resolved, Mn$_{12}$-$t$BuAc
allows the study in unprecedented detail of the crossover between
thermally assisted and pure quantum tunneling.
The dominant field steps for each step index $n$
are shown in Fig.~\ref{H_step} by filled circles.
Whereas the crossover is gradual for $n$ = 9 and 10,
a clear step is seen for $n$ = 8. Indeed, the quantum
step for (8:1) is for all temperatures smaller
than either (8:0) or (8:2). Similar results
are found for $n$ = 7 and 6. For $n \leq 5$, the
crossover goes directly from non-measurable steps to
finite ones with $p \geq 3$. Note that this
sharp transition has not been observed in 
Mn$_{12}$-Ac~\cite{Kent00,Kent00b,ChiorescuMn12PRL00}.
It has been predicted that a sharp crossover can
be smoothed out by applying a transverse field~\cite{Garanin02}.
This can indeed be observed in Mn$_{12}$-$t$BuAc.
Fig.~\ref{hyst_T_h_tr} shows that about 0.14 T sufficiently
increases the tunnel rate to smooth out the transition.
Similar results were found for $n<8$.

In conclusion, resonance tunneling measurements on a new 
high symmetry Mn$_{12}$-$t$BuAc molecular nanomagnet show 
levels of detail not possible with Mn$_{12}$-Ac, as 
a result of the much less disorder and higher symmetry 
in the crystals of the former. This has permitted an 
unprecedented level of analysis of the data to be 
accomplished, resulting in information not attainable 
with Mn$_{12}$-Ac. The crossover between thermally 
assisted and pure quantum tunneling can be easily 
explored and is found to be abrupt or gradual depending 
on the magnitude and orientation of the applied field. 
Simulation of the data allows  $D$ and $B$ to be 
directly obtained from the step fine structure for the first time.



\begin{thebibliography}{10}
\bibitem{note1}
The field $H_z$ seen by a spin is roughly
the sum of the applied field $H_{\rm appl}$ and the demagnetization 
field of the crystal. The latter can be determined
by sweeping the applied field back and forth over
a resonance transition. We fould a nearly linear
dependence which yield the following field correction:
$\mu_0H_z = \mu_0H_{\rm appl} + 0.04 {\rm ~T}~M/M_{\rm S}$
\bibitem{note2}
The studied crystal contained about 5\% of molecules which
showed faster relaxation, similar to crystals of Mn$_{12}$-Ac.
The corresponding field steps of the impurity phase were ignored.
\end{thebibliography}

\end{document}